\documentclass[epj,nopacs]{svjour}

\usepackage{amsmath,amssymb,amsfonts}
\usepackage[export]{adjustbox}
\usepackage{booktabs}
\usepackage[hidelinks]{hyperref}
\usepackage{color}
\hypersetup{
   colorlinks,
   linkcolor={blue},
   citecolor={blue},
   urlcolor={blue}
}
\usepackage{cite}
\usepackage[shortlabels]{enumitem}
\usepackage[utf8]{inputenc}
\usepackage{dsfont}

\newcommand{\tr}{\mathrm{tr}}
\newcommand{\re}{\mathrm{Re}}

\newcommand{\e}{\mathrm{e}}
\renewcommand{\d}{\mathrm{d}}
\renewcommand{\eqref}[1]{Eq.~(\ref{#1})}
\newcommand{\figref}[1]{Fig.~\ref{#1}}

\def\rhocrit{\rho_{\mathrm{crit}}}
\def\Qimp{Q_{\mathrm{imp}}}

\let\originalleft\left
\let\originalright\right
\renewcommand{\left}{\mathopen{}\mathclose\bgroup\originalleft}
\renewcommand{\right}{\aftergroup\egroup\originalright}

\begin{document}

\title{\boldmath Estimating $\chi_\mathrm{top}$ Lattice Artifacts from Flowed SU(2) Calorons}

\journalname{Eur. Phys. J. C  (2019) 79:510}

\author{P. Thomas Jahn\inst{1}\and Guy D. Moore\inst{1}\and Daniel Robaina\inst{1,2}}
\institute{Institut f\"ur Kernphysik (Theoriezentrum), Technische Universit\"at Darmstadt\\ Schlossgartenstra{\ss}e 2, D-64289 Darmstadt, Germany\and Max-Planck-Institut f\"ur Quantenoptik\\ Hans-Kopfermann-Stra{\ss}e 1, 85748 Garching, Germany}

\date{Received: 14 January 2019 / Accepted: 1 June 2019}

\abstract{
Lattice computations of the high-temperature topological
susceptibility of QCD receive lattice-spacing corrections and
suffer from systematics arising from the type and depth of gradient
flow.  We study the lattice spacing corrections to $\chi_\mathrm{top}$
semi-analytically by exploring the behavior of discretized
Harrington-Shepard calorons under the action of different forms of
gradient flow. From our study we conclude that $N_\tau = 6$ is
definitely too small of a time extent to study the theory at
temperatures of order $4~T_\mathrm{c}$ and we explore how the
amount of gradient flow influences the continuum extrapolation.
}

\maketitle

\section{Introduction \label{sec:Introduction}}
Quantum Chromodynamics fields generically contain topological
structure, which may play a role in some low-energy phenomenology and
have been treated extensively in the literature
\cite{Schafer:1996wv}.  This paper shall put the focus on topology at
high temperatures $T \gg T_{\mathrm{c}}$ the (crossover) temperature.
At these temperatures, the topological susceptibility -- the mean-squared
$Q$-value per unit 4-volume -- is small, but it nevertheless plays
a crucial role in the cosmological physics of the QCD axion
\cite{Peccei:1977ur,Weinberg:1977ma,Wilczek:1977pj,Preskill:1982cy,Abbott:1982af,Dine:1982ah}, should it exist.
In this case, besides the now rather well-determined
value of the topological susceptibility in
vacuum \cite{diCortona:2015ldu}, we also need the topological
susceptibility in the temperature range $3~T_\mathrm c$ -- $7~T_\mathrm c$
\cite{Klaer:2017ond,Moore:2017ond}, since this is the temperature
range where a cosmological axion density would be established.  The
exploration of topological susceptibility at high temperatures within
QCD has recently become the topic of intense investigation
\cite{Berkowitz:2015aua,Borsanyi:2015cka,Petreczky:2016vrs,Taniguchi:2016tjc,Burger:2017xkz,Frison:2016vuc,Borsanyi:2016ksw,Jahn:2018dke}.
(For a recent review of the axion see Ref.~\cite{Irastorza:2018dyq}.)

At low temperatures, topology is dominated by long range physics where
the coupling is strong. In this regime, we have little rigorous theoretical
guidance for the form of the configurations which predominantly
contribute to the topological susceptibility.
On the other hand, at high temperatures the coupling is weak and 
Gross, Pisarski, and Yaffe \cite{Gross:1980br} demonstrated that
the dominant configurations are expected to be calorons (periodic
instantons) with radius $\rho$ of order of half the inverse
temperature. Smaller objects are suppressed by the smaller coupling
at short distances, and larger objects are suppressed by their interaction
with thermal fluctuations. In general, at high
temperatures, topological objects are rare and can be treated singly
(dilute instanton gas).

At the temperature range relevant for axion cosmology, topological
objects should be dilute, but corrections to perturbation theory may
be large. In particular, the calculation of
the thermal corrections to the caloron action is not secure due to the
infrared issues of thermal fluctuations which exist even at weak
coupling in gauge theories \cite{Linde:1980ts,diCortona:2015ldu}. Those can influence the
environment of the caloron and change the exact coefficient of the
$\exp(-\rho^2 T^2)$ suppression of large calorons. Therefore a
nonperturbative lattice treatment would be very valuable.  But this
brings with it some complications.
Strictly speaking, there is no perfectly clean and unambiguous
definition for topology on the lattice. After all, on a compact space
without boundary, topology partitions the space of smooth continuum
configurations into path-disconnected subspaces with different integer
$Q$ values. But the space of lattice configurations is
$[\mathrm{SU}(N)]^{N_\ell}$
(with $N_\ell$ the number of lattice links), which is path connected,
precluding a continuous and integer definition of $Q$. This problem
has long been appreciated; L\"uscher showed that
topology becomes well defined if we restrict to sufficiently
``smooth'' lattice configurations, in the sense that all plaquettes
are suitably close to the identity \cite{Luscher:1981zq}. That means that the failure for a
perfectly clean definition of topology on the lattice lies with
certain very non-smooth configurations, termed ``dislocations.''%
\footnote{Of course topology can be given a rigorous definition, for
  instance, by the signed sum of zero modes of a Ginsparg-Wilson
  \cite{Ginsparg:1981bj}
  fermionic operator.  However this definition is not unique, since
  there are multiple choices for Ginsparg-Wilson operators.  For
  instance, those operators constructed by the overlap method
  \cite{Neuberger:1997fp,Luscher:1998pqa}
  are dependent on specific choices in the Wilson operator used to
  build the overlap operator.  Roughly speaking, dislocations are
  those configurations which will give different values of topology
  for different implementation choices.}

A modern tool for studying topology on the lattice is the integration
of a lattice discretized version of $Q$ after the lattice fields have
been subjected to gradient flow
\cite{Narayanan:2006rf,Luscher:2009eq}. This definition has its roots
in older studies employing \textsl{cooling}
\cite{Berg:1981nw,deForcrand:1997esx,GarciaPerez:1998ru,GarciaPerez:1999hs},
with gradient flow representing a better-controlled and
better-understood form of gauge-link cooling. With gradient flow, a
well-defined parameter $t$
controls the extent of smearing applied.
Gradient flow tends to eliminate dislocations \cite{Teper:1985rb},
but since there is no clean distinction between dislocations and
small-but-physical instantons, it may also destroy the smallest
instantons which we want to keep.  This issue is addressed
phenomenologically in most lattice studies that attempt a calculation
of the topological susceptibility, but we are surprised by the
absence of a more systematic study, which might help in understanding
how wide the flow time window actually is, and how
large we should expect the lattice artifacts in the topological
susceptibility to be.  In other words, it would be useful to get a
better analytical understanding of how lattice spacing and flow-depth
effects may influence the determined topological susceptibility.

In this paper we shall explore this issue by studying exactly how much
gradient flow destroys exactly what size of
Harrington-Shepard calorons \cite{Harrington:1978ve}. This requires a
lattice implementation of the caloron, which we supply. We also
explore different implementations of gradient flow: a.) Wilson flow
\cite{Luscher:2009eq}, b.) a recently proposed
$\mathcal{O}\left(a^2\right)$-improved flow dubbed Zeuthen flow
\cite{Ramos:2015baa}, and c.) an ``overimproved'' flow in which we force
the $a^2$ errors in the flow action to have the opposite sign as in
Wilson flow. We are hardly the first to implement discretized
topological configurations on the lattice
\cite{Pugh:1989ek,Bruckmann:2004nu,Bruckmann:2004zy} or to consider
topology after cooling \cite{deForcrand:1995qq,deForcrand:1995bq,deForcrand:1997esx,GarciaPerez:1998ru,GarciaPerez:1999hs}. But
our emphasis is a little different; we want to understand and control
what size of dislocation/caloron survives what amount of flow, and
what impact this may have on the determination of the topological
susceptibility at finite lattice spacing and on the corresponding
extrapolation to the continuum limit.

A natural objection to our study is that, in the temperature range of
relevance, the coupling is still quite large.  Therefore,
the actual topological objects from the lattice will not
be clean calorons, but will have large fluctuations.  Nonetheless, gradient
flow will drive any topological object towards a caloron solution in a
much smaller amount of flow time than it takes for the object to
disappear, since the caloron is a stationary point of
the action up to $a^2$ corrections.  Therefore the actual topological
objects' flow trajectories should be very similar to those for
calorons.  (In fact, in the continuum, we could even define the size
of a topological object to be the size of the caloron it approaches
under gradient flow.) Therefore our study can still shed light on how
much flow removes what size of topological object.  Combining this
with an estimate, based on Gross Pisarski and Yaffe's work, for the
size distribution of calorons, can illuminate what flow depths affect
the topological objects we want to keep, and what lattice spacings may
be too coarse to distinguish between dislocations and physically
relevant topological objects.

The paper is structured as follows: In Sec.~\ref{sec:discretization}
we collect our definitions and define our topology configurations at
finite temperature. Section~\ref{sec:gradientflow} then introduces the
different flows and flow actions. In Sec.~\ref{sec:results} we compute
$\rhocrit(t)$ which indicate the value of the radius of a caloron that
barely survived the amount of flow time $t$. These curves are further
used in Sec.~\ref{sec:application} to estimate how lattice spacing
systematics interfere with flow effects in the approach to the
continuum. An example study at $T=4~T_\mathrm{c}$ is used. Our
conclusions can then be found in Sec.~\ref{sec:conclusion}.

\section{\boldmath Caloron discretization with $Q=1$ \label{sec:discretization}}

The Harrington-Shepard caloron \cite{Harrington:1978ve} can be
understood as the finite temperature generalization of the BPST
instanton \cite{tHooft:1976rip}. The most straightforward way of constructing the solution
is to recognize that one has to take into account the infinite time
copies that arise due to the compactification of the Euclidean time
direction whose inverse length plays the role of temperature ($\beta
\equiv T^{-1}$). Following \cite{Jackiw:1976fs} we consider a superpotential of the form 
\begin{align}
  \label{eq:caloronpotential}
  \Phi_\mathrm{HS}(x) &= 1 + \sum_{k \in \mathds{Z}}
  \frac{\rho^2}{\left( x-z_k \right)^2} \nonumber \\
  &= 1 + \frac{\pi \rho^2 \sinh
    \frac{2 \pi \left| \vec x - \vec z \right|}{\beta}}{\beta \left|
    \vec x - \vec z \right| \left( \cosh \frac{2 \pi \left| \vec x -
      \vec z \right|}{\beta} - \cos \frac{2 \pi \left( x_0 -z_0
      \right)}{\beta}\right)}
\end{align}
with $z_k = \left(z_0 + k\beta, \vec{z}\right)$ and $\rho$ having length-units and playing the role of the caloron radius. Concentrating on SU(2) calorons, the gauge field continuum form is obtained via
\begin{equation}
    A_\mu^a(x) = \eta_{a\mu\nu} \partial_\nu \ln \Phi_\mathrm{HS}(x),
    \label{eq:gaugefield}
\end{equation}
where $\eta_{a\mu\nu} = \epsilon_{0a\mu\nu}+\delta_{a\mu}\delta_{\nu0} - \delta_{a\nu}\delta_{\mu0}$
is the 't Hooft symbol (equivalently for $\bar{\eta}_{a\mu \nu}$ the sign of the last two terms is reversed
and one obtains an anti-instanton with $Q=-1$) \cite{tHooft:1976rip}. The
self-duality condition $F_{\mu \nu} = \frac{1}{2}\epsilon_{\mu \nu \rho
\sigma}F_{\rho \sigma}$, implies that a valid solution for $A_\mu(x)$ is
obtained if $\Phi_\mathrm{HS}(x)$ obeys the Poisson equation $\square
\Phi_\mathrm{HS} = 0$.

It is a tedious but straightforward exercise to check that indeed this gauge field
yields $Q=1$ and $S=8\pi^2$. Notice that it is given in singular
gauge, meaning that there is a singularity at the center of the
caloron when $x=z$ \footnote{%
  On the lattice, we will
  avoid the singularity by placing our topology objects in between
  lattice points and unless stated otherwise we always consider
  $z~=~(z_0, \vec{z}) = \frac{1}{2}\left(\beta -a, L-a\right)$.}.
Next, we use the path-ordered exponential map to obtain the expression
for the lattice links as 
\begin{equation}
    U_\mu(x) = \mathcal P \exp \left[a \int_0^1 \d t \ A_\mu\left(\Gamma_\mu(x,t)\right) \right],
    \label{eq:pathordered}
\end{equation}
where $\Gamma_\mu(x,t) = x + t a \hat \mu$ is an appropriate
parameterization for the corresponding path connecting the two
neighboring lattice sites $x$ and $x+a\hat{\mu}$ with $t\in [0,1]$ (no
summation over $\mu$ is implied in \eqref{eq:pathordered}).

In the case of an instanton (without periodic images), $A_\mu$ commutes everywhere along a link,
but this turns out not to be true for the caloron. In order to compute the links we rewrite \eqref{eq:pathordered} as a product of $n$
shorter links,
\begin{equation}
  \label{linklimit}
  U_\mu(x) = \lim_{n\to\infty} \: \mathcal{P} \prod_{k=1}^n \exp \left[ \frac{a}{n}
  A_\mu\left(\Gamma_\mu\left(x,\frac{2k-1}{2n}\right)\right) \right],
\end{equation}
where in practice we use $n \sim 40$ rather than taking the strict
$n \to \infty$ limit. Periodicity is imposed in all four dimensions, and while 
boundary effects are absent in the time direction, our lattice field will have a discontinuity at the spatial edges of the box. We
ameliorate this problem with gradient flow confined to the boundary
region as described in App.~\ref{app:boundaryflow}.

Finally the embedding into an SU(3) background is trivial since a particular
lattice gauge exists in which the links take the following form:
\begin{equation}
  \label{embed}
  U^{\mathrm{SU(3)}}_\mu = \left(\begin{array}{c|c}
    U^{\mathrm{SU(2)}}_\mu & \begin{matrix} 0 \\ 0\end{matrix}\\
    \hline
    \begin{matrix}0 & 0 \end{matrix} & 1 \\
    \end{array}\right).
\end{equation}
Therefore we are \textsl{effectively} considering SU(2) configurations
in this paper. This makes our study less general than
Ref.~\cite{Bruckmann:2004nu}, who also consider calorons in the
background of nontrivial holonomy. Note, however, that if we are
primarily interested in \textsl{high} temperatures, nontrivial
holonomy is not likely to be relevant, since fluctuations create an
effective potential for the Polyakov loop which favors trivial
holonomy \cite{Gross:1980br}.

\section{Gradient Flows}
\label{sec:gradientflow}
Gradient flow and its discretized version on the lattice
\cite{Narayanan:2006rf,Luscher:2009eq,Luscher:2010iy,Luscher:2010we,Luscher:2011bx,Luscher:2013cpa}
have become an essential tool to reduce UV fluctuations.
In the continuum it defines a mapping of the gauge fields
$A_\mu(x)$ to smeared gauge fields $B_\mu(x,t)$, where
$t$ is the so-called flow time, via the flow equation
\begin{equation}
  \label{flowdefs}
  \partial_t B_\mu = D_\nu F_{\nu\mu}, \qquad B_\mu(x,0) = A_\mu(x).
\end{equation}
The right-hand side of this differential equation is nothing but the classical
equations of motion. Consequently, it will drive the gauge field along the
trajectory of steepest descent minimizing the action along the way.
On the lattice the simplest form reads
\begin{equation}
    a^2 \partial_t V_\mu(x,t) = - g_0^2  \partial_{x,\mu} S_\mathrm W [V] V_\mu (x,t),
    \label{eq:Wilson}
\end{equation}
where $V_\mu(x,t)$ is the flowed gauge field with initial condition
$V_\mu(x,0) = U_\mu(x)$ and $S_\mathrm{W}$ denotes the Wilson
plaquette action. The Lie-algebra valued derivative of a general
function of the link variables $f(U_\mu)$ is given by
\begin{equation}
  \label{linkderiv}
    \partial_{x,\mu}^a f(U_\mu(x)) = \left. T^a \frac{\d}{\d s}
    f\left(\e^{sT^a} U_\mu(x)\right) \right|_{s=0}.
\end{equation}
 Recently, an $\mathcal O\left(a^2\right)$ improved version
 of the flow equation was developed in Ref.~\cite{Ramos:2015baa}.
 The so-called Zeuthen flow equation reads
\begin{equation}
    a^2 \partial_t V_\mu(x,t) = -g_0^2 \left( 1 + \frac{a^2}{12}
    \nabla_\mu^\ast \nabla_\mu \right)  \partial_{x,\mu} S_\mathrm{Sym}[V] V_\mu(x,t),
    \label{eq:Zeuthen}
\end{equation}
where $S_\mathrm{Sym}$ is the tree-level improved Symanzik
action \cite{Luscher:1984xn} and the discretized adjoint covariant derivative is given by
\begin{equation}
    \begin{split}
        a \nabla_\mu f(x) &= U_\mu(x) f(x+a\hat\mu) U_\mu^\dagger(x) - f(x),
        \\
        a \nabla_\mu^\ast f(x) &= f(x) - U_\mu^\dagger(x-a\hat\mu) f(x-a\hat\mu) U_\mu(x-a\hat\mu).
    \end{split}
\label{discretederiv}
\end{equation}
The unexpected additional factor
$(1+a^2 \nabla_\mu^\ast \nabla^\mu/12)$ can be
understood as follows. We know that improvement requires replacing
square plaquettes with a linear combination of squares and
rectangles. The Symanzik action does this in the four spacetime directions.
The added term does it in the flow-time direction. Ref.~\cite{Ramos:2015baa}
have proven that this gives a (tree-level) $\mathcal{O}\left(a^2\right)$
improvement of the flow equation.
In addition to these two we investigate a flow equation with an overimproved
action (precise definitions of these actions are given in Subsec.~\ref{subsec:actions})
\begin{equation}
    a^2 \partial_t V_\mu(x,t) = -g_0^2 \left( \partial_{x,\mu} S_\mathrm{OI}[V] \right) V_\mu(x,t).
    \label{eq:overimproved}
\end{equation}
We expect this flow equation to allow for stable topological solutions under flow. Notice that the flow equations Eqs.~(\ref{eq:Wilson}), (\ref{eq:Zeuthen}),
and (\ref{eq:overimproved}) do not depend on $g^2_0$ at all since the actions themselves carry such a factor, too. We therefore omit $g^2_0$ factors when not relevant for the discussion. 

The goal is to study the effect of Eqs.~(\ref{eq:Wilson}), (\ref{eq:Zeuthen}),
and (\ref{eq:overimproved}) on our constructed clean topological
configurations to learn about how their topological properties are changed.
This can represent interesting information to better control systematic
errors when performing lattice calculations of topological observables with the help of flow.

\section{Lattice Caloron Properties}
\label{sec:results}
Mainly we will focus on the measurement of the topological charge $Q$
and the action $S$ which in the continuum take the values $Q=1$ and
$S=8\pi^2$, respectively. Deviations from these numbers occur on a
finite lattice due to cutoff and boundary effects. We will try to
disentangle those to more deeply
understand the effect of flow.

\subsection{Actions}
\label{subsec:actions}

Throughout this work we consider L\"uscher-Weisz actions of the form \cite{Luscher:1984xn,openQCD} (omitting $g^2_0$ factors)
\begin{align}
  \label{generalaction}
  S(c_0,c_1) &= 2 \left(c_0\sum_{x}\re\tr\left(\mathds{1}-U_\mathrm P(x)_{(1,1)}\right)\right.\nonumber\\
  &+ \left.c_1\sum_{x}\re\tr\left(\mathds{1}-U_\mathrm R(x)_{(2,1)}\right)\right),
\end{align}
where $U_\mathrm P(x)_{(1,1)}$ denotes the simple closed plaquette and
$U_\mathrm R(x)_{(2,1)}$ are $2\times1$ (and $1\times2$) rectangles where both
loop orientations are taken into account by the $\re\tr$ operation. A correct normalization requires $c_0 +
8c_1 = 1$. The three actions going into
the three different flow equations can be summarized as follows:
\begin{align}
  S\left(1,0\right) &= S_\mathrm{W} \qquad & \text{(Wilson),} \nonumber \\
  \label{eq:Schoices}
S\left(5/3,-1/12\right) &= S_\mathrm{Sym} \qquad &
\text{(Symanzik),}\\
S\left(7/3,-1/6\right) &= S_\mathrm{OI} \qquad & \text{(Overimproved).}
\nonumber
\end{align}

\begin{figure}[tbh]
    \includegraphics[width=\linewidth]{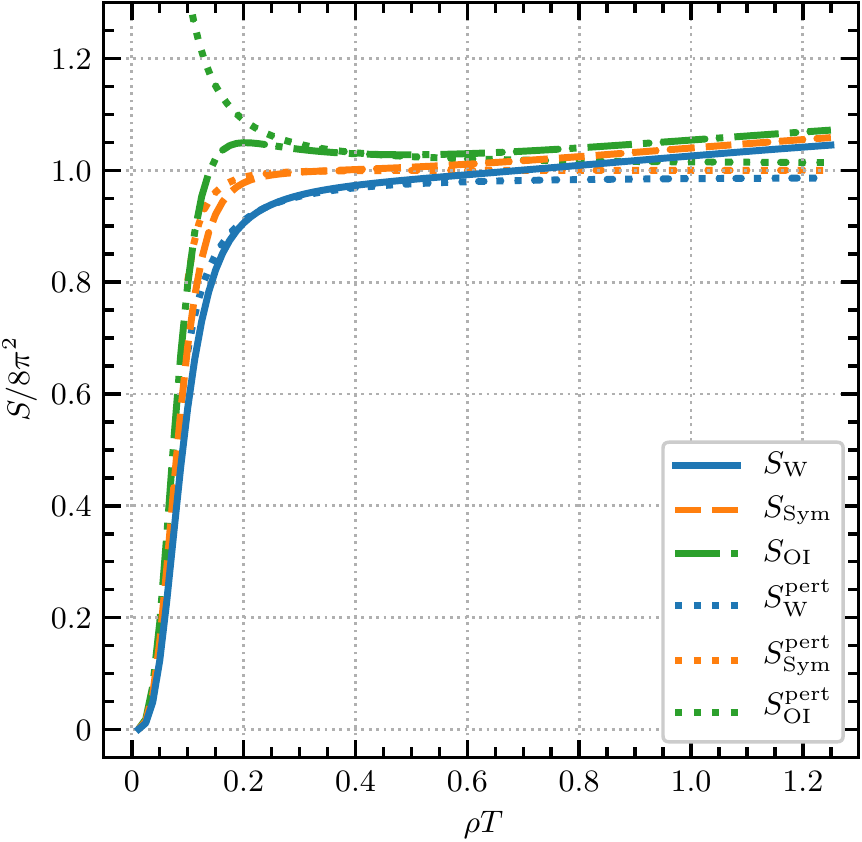}
    \caption{\label{fig:actions_unflowed}Different actions together
      with the perturbative predictions as a function of the caloron
      size. The caloron is placed at the center of an $8\times 48^3$
      lattice. In this plot we used $t_0 T^2=20/64$ for reducing the
      boundary effects (cf. App.~\ref{app:boundaryflow}).
      Dotted lines indicate a small-$a$ expansion up to the first
      nontrivial correction.}
\end{figure}

\figref{fig:actions_unflowed} explores how each action varies as a
function of the caloron size $\rho T$, on a lattice with $N_\tau =8$
and an aspect ratio of 6. We find that all actions start
small and are suppressed until about $\rho=1.6 a$, which is
$\rho T = 0.2$ for this $N_\tau$ value.
A rapid rise towards the expected value of $8\pi^2$ is then observed,
followed eventually by a rise above $8\pi^2$.  These two features --
the rapid rise from zero towards $8\pi^2$, and the eventual rise above
this value -- arise from different artifacts.  The former is a lattice
spacing artifact, which we now explore.
We overlay each curve with an estimate based on a small-$a$ expansion,
taken to first nontrivial subleading order.  Specifically, the
expansion of the Wilson action in operator dimension takes the form
\begin{equation}
S_\mathrm W = -\frac{1}{2}\tr\left\{F_{\mu \nu}F_{\mu
  \nu}\right\} +\frac{a^2}{12}\tr\left\{D_\mu F_{\mu \nu}D_\mu
F_{\mu \nu}\right\} + \cdots,
\end{equation}
where each index is summed once.  This leads to $a^2$ corrections to
the caloron action.  Inserting the caloron field from
Eqs.~(\ref{eq:gaugefield}) and (\ref{eq:caloronpotential}), a
corresponding finite temperature integration
$\int^{1/T}_0 \mathrm d\tau \int \d^3x$ yields
\begin{equation}
S_\mathrm W =
8\pi^2\left[1+\mathcal{F}(\rho T)\left(\frac{a}{\rho}\right)^2 +
  \mathcal{O}\left(\frac{a}{\rho}\right)^4\right].
\label{eq:caloron_pert}
\end{equation}
We have evaluated $\mathcal{F}(\rho T)$ numerically and find that it
is very well fit by
\begin{equation}
\mathcal{F}(\rho T) = -\frac{1}{5} + b (\rho T)^2 + \mathcal{O}(\rho T)^4
\end{equation}
with $-1/5$ the zero-temperature (instanton) value and
$b=-0.758$.  The first (vacuum) effect is $\mathcal{O}(a^2/\rho^2)$; because
of it, the action is significantly smaller for $\rho < 1.6 a$, and we
should consider such objects as ``dislocations'' rather than true
continuum-like calorons.  The second term gives rise to an
$\mathcal{O}(a^2 T^2 = 1/N_\tau^2)$ correction, which is present at all
caloron sizes, and represents a size-independent mis-estimate of the
caloron action due to the lattice spacing.
For the overimproved case the $a^2$ correction is the
same but with opposite sign.  For the Symanzik case we have not
evaluated the full temperature-dependent $\mathcal{O}(a^4)$
correction, but instead use the $\mathcal{O}(a^4)$ correction to the
instanton action found by Ref.\cite{GarciaPerez:1993lic}.  This is
adequate because the correction is small for $\rho T \sim 1$ where
thermal effects are expected.

The overimproved action possesses positive
$a^2$ corrections and therefore develops a maximum. This will be of
importance when flowing with this action as it will stabilize calorons
larger than the size where $S$ is maximum, preventing them from
shrinking, ``falling through the lattice,'' and being lost. As we will
see, although promising, this approach does not substantially help in
the calculation of the topological susceptibility.

The figure also features a rise in the action above $8\pi^2$ at large
caloron sizes; for the lattice considered in the figure, this effect
becomes larger than the $a^2$ effects above about $\rho T = 0.7$.
This is a finite-volume effect which is ameliorated by
going to a larger aspect ratio.  It is also partly an artifact of the
way we construct the caloron solution, since we take properly into
account the temporal periodicity but not the space periodicity.  In
order to reduce this effect as much as possible, we ``flow the boundaries
away.'' What this means is that we perform a space-time dependent flow
where the core of the configuration (where most of the topological
charge is localized) is unaffected while boundary effects are smoothed
out. A similar idea was used in Ref.~\cite{GarciaPerez:1998ru} where a
discretized version of $D_{\mu}F_{\mu \nu}(x)$ was measured on every
space-time point and an improved form of cooling was performed on
those lattice points that satisfied the bound $D_{\mu}F_{\mu \nu}(x) >
\epsilon$. In App.~\ref{app:boundaryflow} we explain our own procedure
for reducing boundary effects which we utilize throughout this work. 
From now on, when setting up a topological configuration, 
we always implicitly apply this procedure. Note that the
gradient flow used to reduce the boundary effects should not be confused
with the usual gradient flow that we apply for some calculations in the
remainder of this work.

\subsection{Topological Charges}
The field-strength tensor $F_{\mu \nu}(x)$ is the main building block for constructing
gauge operators like the topological charge. Apart from the
popular geometrical clover definition (4-plaquette average), we considered an improved
version thereof. To this end, we implement an improved field-strength tensor
$\hat{F}^\mathrm{imp}_{\mu \nu}$ free of $\mathcal{O}\left(a^2\right)$ errors by
considering weighted averages of $1\times1$ plaquettes and $2\times1$
rectangles \cite{Moore:1996wn,BilsonThompson:2002jk}.

\begin{figure}
    \includegraphics[width=\linewidth]{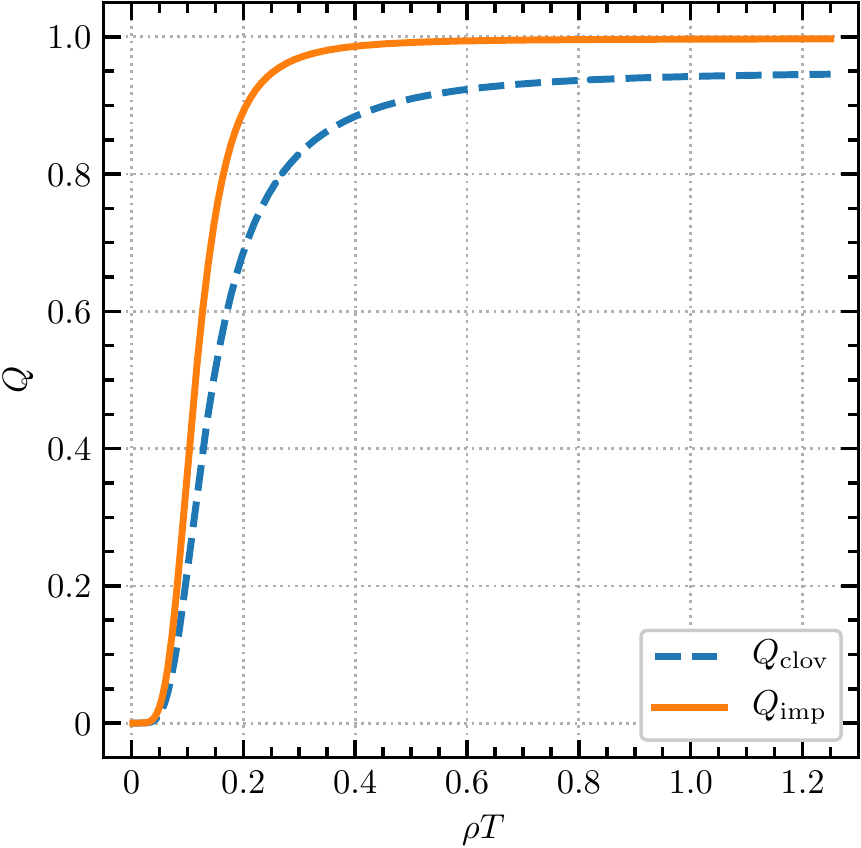}
    \caption{\label{fig:Q_unflowed2}Two different discretizations of the topological charge
      operator as a function of the caloron size on an $8\times 32^3$ lattice.}
\end{figure}

We then study the two definitions
\begin{equation}
  \label{Qdef}
    Q_{\mathrm{clov/imp}} = -\frac{1}{32 \pi^2} \epsilon_{\mu\nu\rho\sigma}\sum_{x} \tr \left( \hat{F}_{\mu\nu}(x) \hat{F}_{\rho\sigma}(x) \right),
\end{equation}
where in each case
\begin{align}
  \begin{split}
  \hat{F}^\mathrm{clov}_{\mu \nu}(x) &= F_{\mu \nu}(x) + \mathcal{O}\left(a^2\right),
    \\
    \hat{F}^\mathrm{imp}_{\mu \nu}(x) &= F_{\mu \nu}(x) + \mathcal{O}\left(a^4\right).
    \end{split}
\end{align} 
As can be seen from \figref{fig:Q_unflowed2}, the improved topological
charge operator shows a much better behavior at all investigated
values of the radius. Notice that boundary effects are milder for the
topological charge than for the action. We see only advantages to
using the improved definition.

\subsection{Critical Radius}
\label{subsec:criticalradius}
\begin{figure}[ht]
    \includegraphics[width=\linewidth]{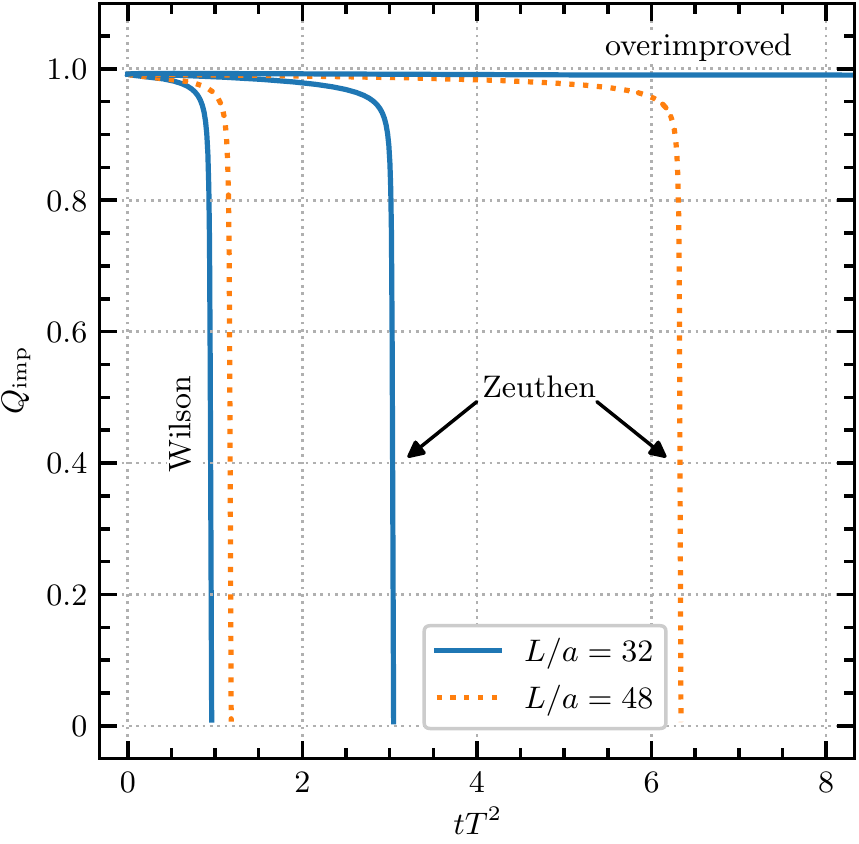}
    \caption{\label{fig:flowuntildeath}Topological charge $\Qimp$ as a function of flow time for
    a caloron with $\rho T=0.5$, for three flow definitions and two box
    sizes ($N_\tau = 8$ is fixed). Calorons live much longer under Zeuthen flow, and forever
    under overimproved flow.}
\end{figure}

One of the relevant aspects we want to address in this paper is the
behavior of a discretized caloron configuration under different flow
equations.  We see in Fig.~\ref{fig:Q_unflowed2} that, for $N_\tau=8$,
a caloron with $\rho T \gtrsim 0.12$ will have a $Q$ value
larger than $1/2$, if the measurement is made before any flow is
applied.  Of course in practice such a measurement would be
impossible, since the $Q$ of the caloron would be swamped by
contributions from nontopological fluctuations.  These disappear after
a modest amount of gadient flow.  But calorons also shrink and tend to
disappear as flow is applied, precisely because of the action
corrections which we explored in Fig.~\ref{fig:actions_unflowed}.
We illustrate this effect in Fig.~\ref{fig:flowuntildeath}, which
shows how the $Q$ value changes under flow for a ``large'' caloron
with $\rho T = 0.5$ on an $N_\tau=8$ lattice.  We see that, after some
amount of time, the measured topological charge abruptly collapses.
This occurs because flow causes the caloron to shrink, eventually
abruptly shrinking away and disappearing between lattice sites.  At
least for Wilson and Zeuthen flow, any caloron will eventually
disappear in this way.
The collapse is slower for Zeuthen flow because of the absence of
$a^2$ lattice-spacing corrections to the action, but the $a^4$ and
finite-volume effects nevertheless eventually lead to collapse.  For
the overimproved action, the action has a maximum which prevents flow
from ever destroying the caloron.

It would be very useful to know more precisely, how much flow destroys
what size of caloron.  To investigate this, we first establish a
definition of when we consider a caloron to really exist:  when the
topological charge $Q$ exceeds some threshold, which we choose to be
$1/2$.  (We see in Fig.~\ref{fig:flowuntildeath} that the exact choice is
almost immaterial.)  This corresponds well to the typical procedure
one will use in determining the topological susceptibility in a
simulation:  a configuration is generated, $Q$ is measured, and then
its value is thresholded to an integer.  The choice $1/2$ corresponds
to thresholding to the nearest integer.
We therefore define the critical radius of
a caloron where it becomes topological as
\begin{equation}
  \label{Qcrit}
  \Qimp\left(\rhocrit\right) \equiv 0.5.  
\end{equation}
\begin{figure*}
    \includegraphics[width=\textwidth]{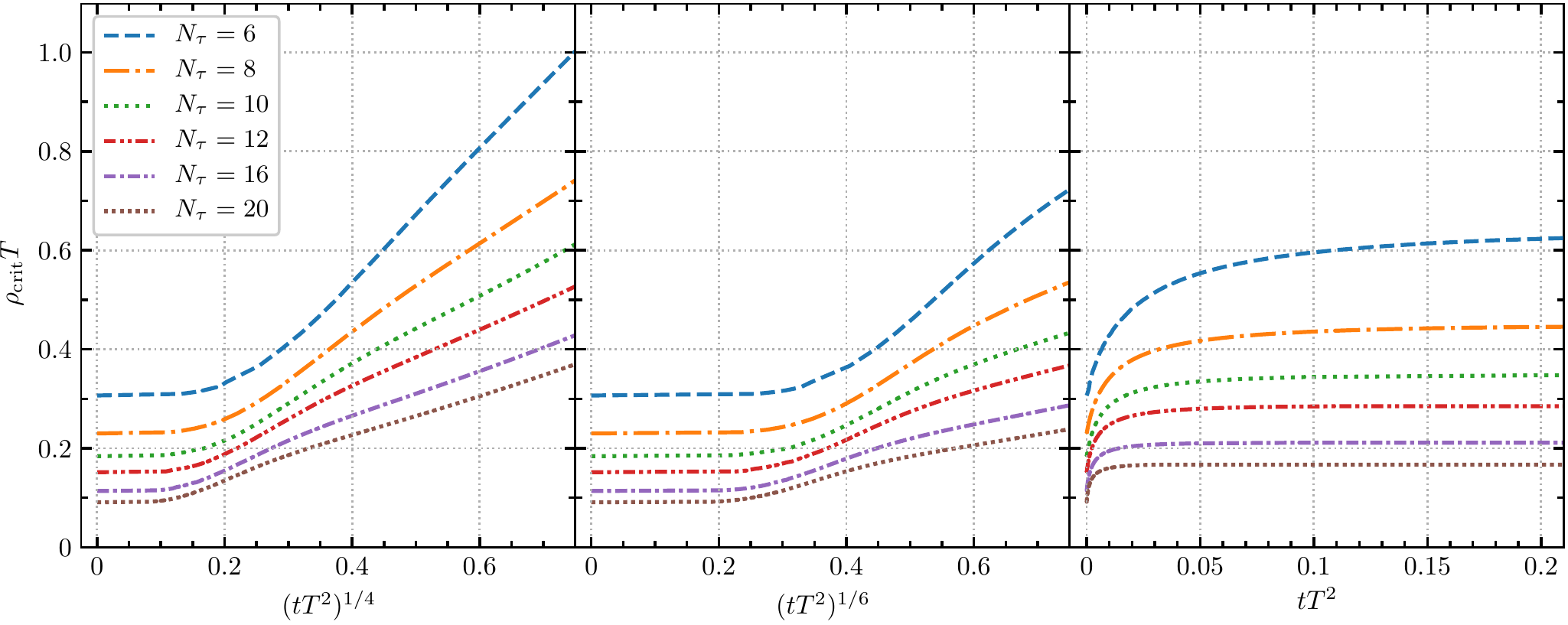}
    \caption{\label{fig:rhocrit}Critical caloron radius (separating calorons which
      survive from those which collapse) as a function of flow depth
      for three types of flow, and several different temporal lattice extents.
      Left: Wilson flow. Middle: Zeuthen flow. Right: Overimproved flow.
      }
\end{figure*}
We can then study how flow causes lattice calorons to shrink and disappear by
investigating $\rhocrit$ as a function of flow time, that
is, what initial caloron radii $\rho$ still have
$\Qimp > 0.5$ after some flow depth $t$.
This is shown in
\figref{fig:rhocrit}, which can be used to look up how much flow
is needed to collapse calorons of a given size.

We see in the figure that $\rhocrit$ grows almost linearly with
$t^{1/4}$, which is easily explained analytically.
An ordinary perturbative fluctuation of momentum $p$
decays as $\exp\left(-p^2 t\right)$ (this is a leading result coming
from \eqref{flowdefs}), and so doubling the size requires four times
the flow time, or $t \propto \rho^2$. However, calorons are nearly
extrema of the action, up to $a^2/\rho^2$ corrections in the Wilson
action, so we expect $t \propto \rhocrit^4$. Therefore
\figref{fig:rhocrit} plots $\rhocrit$ against $t^{1/4}$ (Wilson case), which
would be a straight line for $1/N_\tau \ll \rho T \ll 1$. The figure shows
that calorons also collapse under Zeuthen flow, though more slowly (as
the energy depends on scale only at $\mathcal{O}\left(a^4\right)$, we
therefore expect $t \propto \rhocrit^6$) and, in fact, the curves show a linear trend. Under overimproved flow calorons are preserved above
some critical size, which was the original motivation for considering it \cite{GarciaPerez:1993lic}.

\section{\boldmath Estimated $a^2$ Errors in the Topological Susceptibility}
\label{sec:application}
We want to apply our results to get a semi-analytical understanding of
how both flow depth and $a^2$ errors may influence lattice determinations of
the topological susceptibility at high temperatures.  Our goal is
\textsl{not} to calculate the topological susceptibility \textsl{per
  se,} but to see how flow and lattice spacing may influence its
determination at finite lattice spacing.

\begin{figure}
	\includegraphics[width=\linewidth]{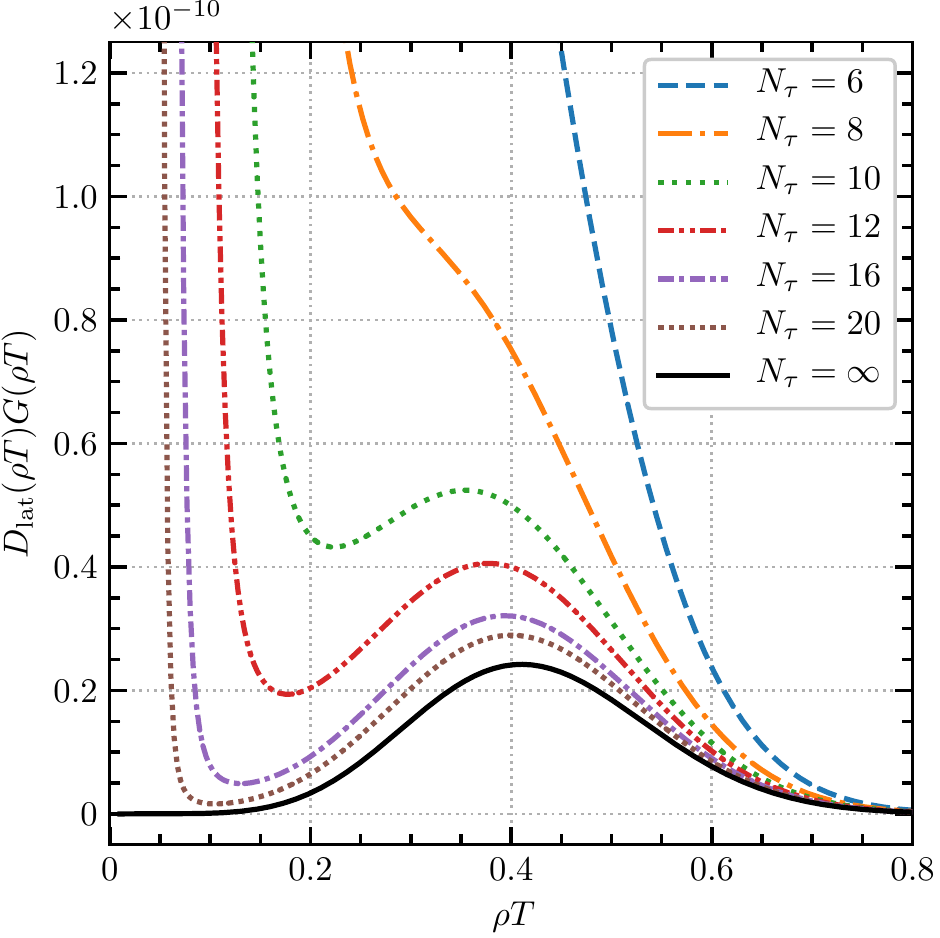}
	\caption{\label{fig:DlatG} Integrand of \eqref{eq:chilat} at
$T/T_\mathrm c=4$ for $N_\tau=6,8,10,12,16,20$. The black solid curve represents
the continuum case and corresponds to the intregrand of \eqref{eq:chicont}.}
\end{figure}

We will do so by approximating the distribution of topological objects
using the dilute instanton gas (DIGA) approximation. We
incorporate the known one-loop renormalization
contributions to the caloron \cite{Gross:1980br}, and
estimate the topological susceptibility by integrating over all
instanton sizes. In the continuum this quantity is given by
\begin{equation}
\chi\left(T/T_\mathrm c \right) \simeq 2 \int^{1/\Lambda^{N_\mathrm f=0}_{\overline{\mathrm{MS}}}}_{0} \d\rho\,
D(\rho) G(\pi \rho T)
\label{eq:chicont}
\end{equation}
with 
\begin{equation}
  \label{Ddef}
  D(\rho) = \frac{d_{\overline{\mathrm{MS}}}}{\rho^5}
  \left(\frac{8\pi^2}{g^2(\mu=\rho^{-1})}\right)^6
  \exp\left(-\frac{8\pi^2}{g^2(\mu=\rho^{-1})}\right)
\end{equation}
the vacuum density of instantons with size $\rho$, and
\begin{align}
  \label{Gdef}
  G(\lambda) &= \exp\left(-2\lambda^2 - 18A(\lambda)\right),\\
  \label{Adef}
  A(\lambda) &= -\frac{1}{12}\ln\left(1+\frac{\lambda^2}{3}\right)
+ \alpha\left(1+\gamma \lambda^{-\frac 32}\right)^{-8}
\end{align}
the thermal corrections, first computed by Gross, Pisarski, and Yaffe
\cite{Gross:1980br}. The parameter values in these expressions are
$\alpha = 0.0128974$, $\gamma = 0.15858$, and
$d_{\overline{\mathrm{MS}}} = \frac{\e^{5/6}}{\pi^2}\e^{-4.534122}$.
The running of the coupling
$g^2(\mu)$ can be found in Ref.~\cite{Chetyrkin:1997un} and
$T_\mathrm{c}/\Lambda^{N_\mathrm f=0}_{\overline{\mathrm{MS}}} = 1.26$
is taken from Ref.~\cite{Borsanyi:2012ve}. The product of $D(\rho)$ and
of $G(\pi \rho T)$ in \eqref{eq:chicont} leads to an integrand with a
broad peak near $\rho  T \simeq 0.4$ (solid black curve in
\figref{fig:DlatG}), which is then the typical size for the
calorons which dominate the topological susceptibility.

In performing a lattice Monte-Carlo study, the practitioner chooses an
action for sampling configurations.  The choice is logically
independent from the choice of action used in gradient flow, but it
can be equally impactful.  In particular, if the lattice study is
based on sampling with the Wilson action, something we assume throughout this section, then the continuum action
$8\pi^2$ in \eqref{Ddef} should be replaced by the lattice Wilson
action of a caloron from \eqref{eq:caloron_pert}, leading to an $a^2$
correction to $D(\rho)$:
\begin{align}
  \label{Dlatt}
  & D_\mathrm{lat}(\rho,T/T_\mathrm c,N_\tau) = \nonumber\\
 & D(\rho)
  \exp\left[-\frac{8\pi^2}{g^2(\mu=\rho^{-1})}
  \left(\frac{1}{\rho T N_\tau}\right)^2
  \mathcal{F}\left(\rho T\right)\right].
\end{align}
This rests on an assumption that the coefficient in front
of the dimension-6 $a^2$-suppressed operator ($DFDF$) takes its tree
level value.  Realistically we expect corrections from, e.g., the
renormalization of the $a^2$ action correction and from higher-order
effects in $G(\pi \rho T)$, so our results here should be viewed only
as estimates, based on the best tools we have available, for how $a^2$
effects will affect the caloron density on the lattice.

To estimate the topological susceptibility as measured on the lattice,
we integrate this modified caloron density over those caloron sizes
which are not destroyed by gradient flow -- which is precisely all
$\rho > \rhocrit$ as determined in \figref{fig:rhocrit}.
We therefore write
\begin{align}
&\chi_\mathrm{lat}(T/T_\mathrm c, N_\tau, tT^2) = \nonumber \\
&2 \int^{1/\Lambda^{N_\mathrm f=0}_{\overline{\mathrm{MS}}}}_{\rhocrit(tT^2, N_\tau)}
\d\rho\, D_\mathrm{lat}\left(\mu =\rho^{-1},T/T_\mathrm c\right) G(\pi \rho T).
\label{eq:chilat}
\end{align}
We illustrate the integrand for several $N_\tau$ values in
\figref{fig:DlatG}.

\begin{figure*}[t]
\includegraphics[width=\textwidth]{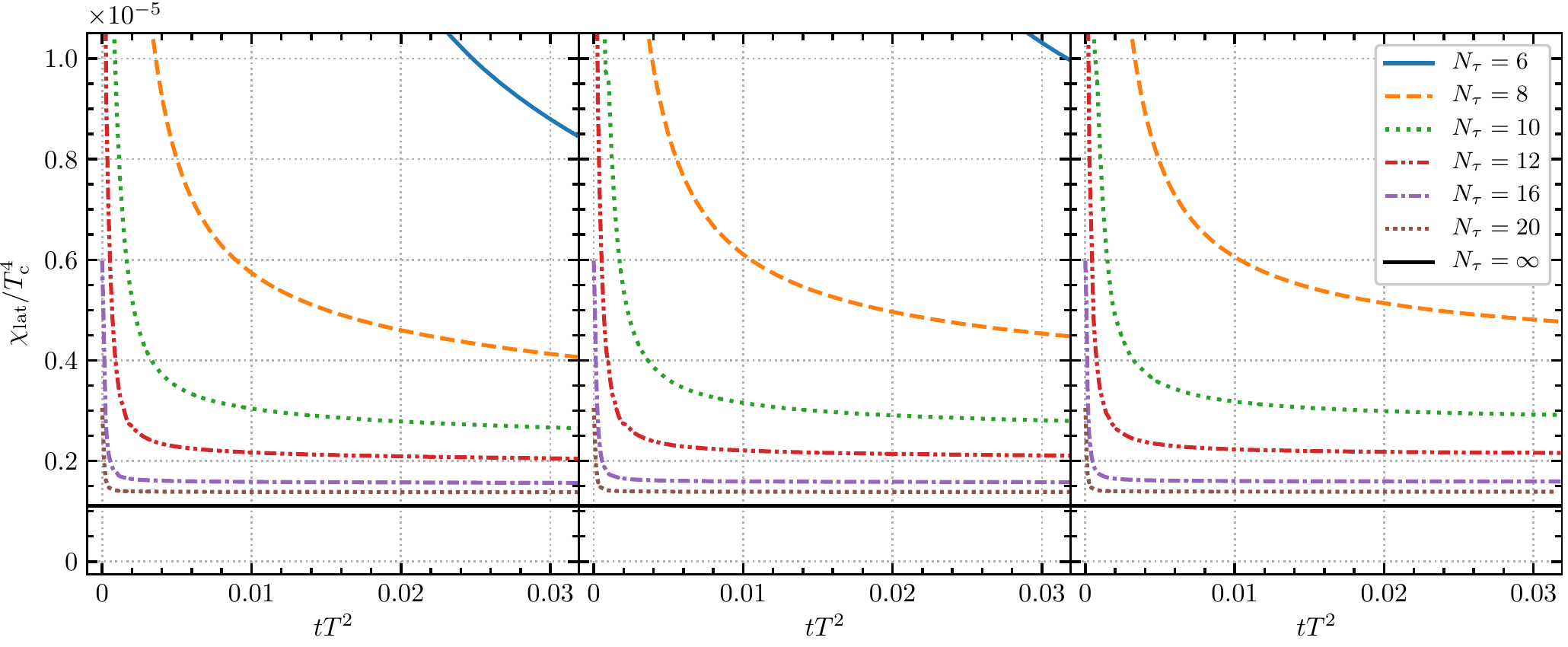}
\caption{\label{fig:chilat}Estimated effect of lattice artifacts on the topological
  susceptibility, at $T/T_\mathrm c=4$ and for $N_\tau=6,8,10,12,16,20$, as
  a function of the flow depth $tT^2$. Left: Wilson flow.
  Middle: Zeuthen flow. Right: Overimproved flow.
  The solid line denotes the continuum limit.}
\end{figure*}
\figref{fig:chilat} shows the resulting estimate of the
topological susceptibility which we would obtain by working at a given
$N_\tau$ and applying a given amount of gradient flow. The lattice
corrections to the action raise the contributions in the peak of \eqref{eq:chicont}
near $\rho T \simeq 0.4$. But lattice artifacts also dramatically increase the
number of dislocations with $\rho \sim a$, as we see from the
integrand in \figref{fig:DlatG}. If these two scales, $a$
and $0.4/T$, are well separated, then gradient flow can erase the
dislocations with little impact on the typical calorons.  That is,
there will be a minimum in the integrand of \figref{fig:DlatG}, and we
can use \figref{fig:rhocrit} to choose a flow depth which will erase calorons
below this minimum.  This leads to a plateau in the susceptibility
over a range of flow depths, as seen in \figref{fig:chilat}.
For coarser lattices such as $N_\tau=6,8$, examining
\figref{fig:DlatG}, it is not clear where to cut to separate calorons
from dislocations, and there is no associated plateau in
\figref{fig:chilat}.  Therefore $N_\tau=6, 8$ will likely not be
sufficient to give results which are stable against the
amount of flow, but larger $N_\tau$ will, especially if we use Zeuthen
flow. Overimproved flow is good for completely ``cleaning'' a
configuration of perturbative fluctuations, but in terms of
eliminating small instantons, it is effectively equivalent to using a
specific fixed depth of Wilson flow. Therefore it is not preferred if we want
\textsl{flexibility} in choosing the size of caloron/dislocation which
we eliminate.
\begin{figure*}[t]
\includegraphics[width=\linewidth]{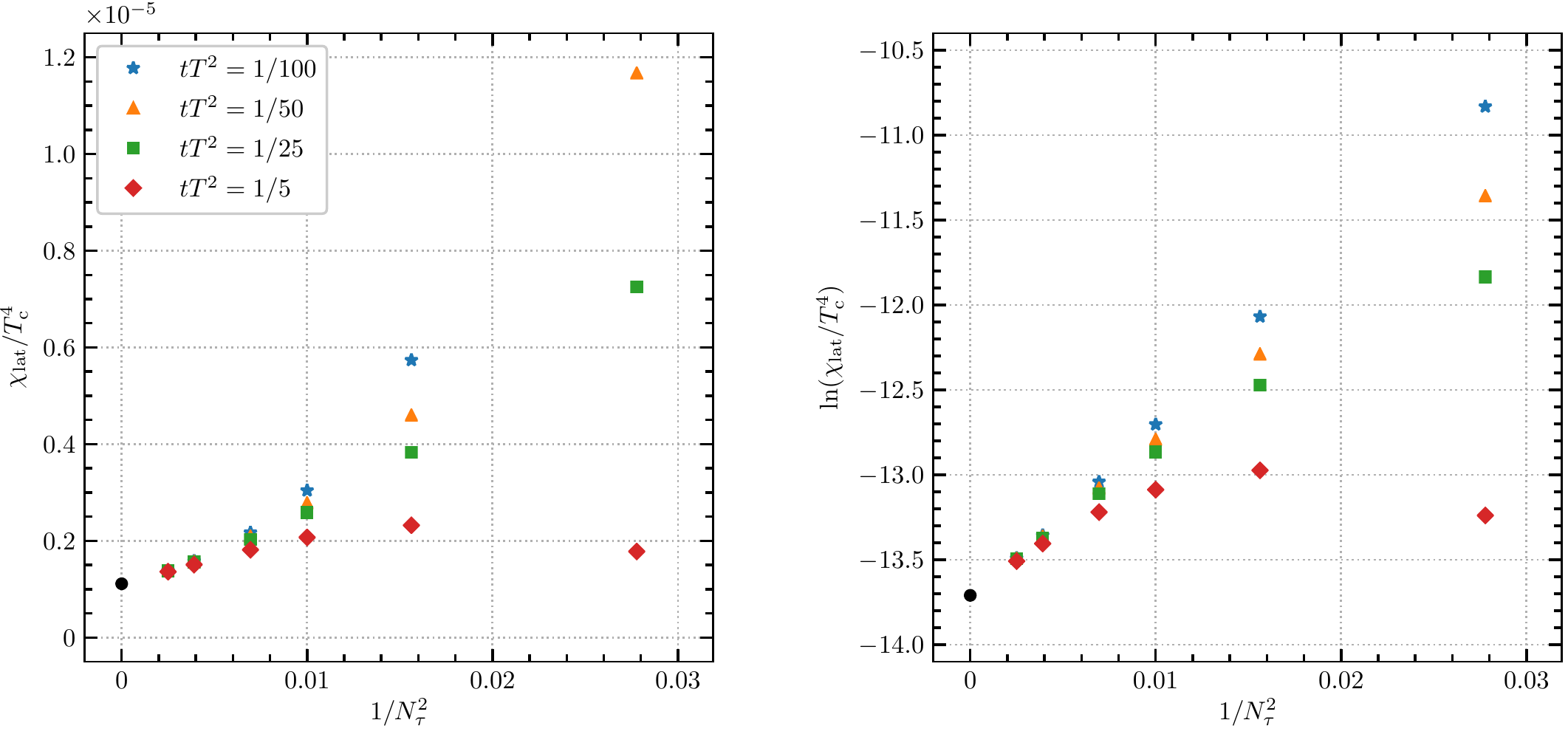}
\caption{\label{fig:chilat_cont}Left: Continuum extrapolation \eqref{eq:chilat} at $T/T_\mathrm{c}=4$
at fixed (Wilson) flow time.
Right: Logarithm of left plot. The black point represents the continuum value.}
\end{figure*}

Finally, we consider the extrapolation to zero lattice spacing in
\figref{fig:chilat_cont}. The lattice spacing corrections are
very large even for $N_\tau = 10$, and a simple extrapolation in
$\chi$ can easily lead to a negative result. But that is because the
$a^2$ errors are best viewed as a correction to the \textsl{logarithm}
of $\chi$, as we see in \eqref{Dlatt}. If we extrapolate in terms of
$\ln(\chi)$, the procedure works much better.

\section{Conclusions}
\label{sec:conclusion}
We have constructed calorons on the lattice. We find that they
possess most of the action and topological charge
if $ \rho T  \gtrsim 1/N_\tau$, and nearly all of the charge
and action if $\rho T \gtrsim 2/N_\tau$. Wilson flow destroys small calorons, with
progressively more flow destroying larger calorons, as expected.  Our
work expresses this in a quantitative fashion, in \figref{fig:rhocrit}.

Also, if a lattice study were to flow each topologically nontrivial configuration until it loses topological character ($\Qimp<0.5$), and keep track of the
distribution of flow depths needed, then a plot as the one in \figref{fig:flowuntildeath} could be used to turn this result
into a size distribution of the topological objects observed on the lattice.

Using our results to estimate the $a^2$ errors which arise when
computing the topological susceptibility $\chi(T)$ on the lattice, we
find that $N_\tau = 6$ is insufficient to be in the scaling
regime (probably $N_\tau = 8$ as well), and lattice spacing errors are expected to lead to a severe
overestimate of $\chi(T)$ at finite $a$, which may lead to negative
values if we extrapolate $\chi(T)$ against $a^2$. It is more natural
to extrapolate $\ln\left(\chi(T)\right)$ against $a^2$, because this corresponds
better to the way in which $a^2$ errors enter in the susceptibility.

Note that the inclusion of light quarks in \eqref{Gdef} would change
the factor $-2\lambda^2$ to $-(2+N_\mathrm f/3) \lambda^2$, which makes the
dominant size of calorons smaller. Therefore, since $\rho T N_\tau$ becomes
smaller, the corrections in \eqref{Dlatt} become larger (since $\mathcal{F}$ is negative), and the value
of $N_\tau$ needed to reach scaling will be still larger.

\begin{acknowledgement}
The authors acknowledge support by the Deutsche Forschungsgemeinschaft
(DFG) through the grant CRC-TR 211 ``Strong-interaction matter under
extreme conditions.'' We also thank the GSI Helmholtzzentrum and the TU
Darmstadt and its Institut f\"ur Kernphysik for supporting this
research. This work was performed using the framework of the publicly
available openQCD-1.6 package \cite{openQCD}. The authors want to
especially thank Marc Wagner and Margarita Garc\'{\i}a-P\'erez for
interesting discussions at early stages of this work.
\end{acknowledgement}

\appendix

\section{Reducing Boundary Effects}
\label{app:boundaryflow}
\begin{figure*}
    \includegraphics[width=1\linewidth]{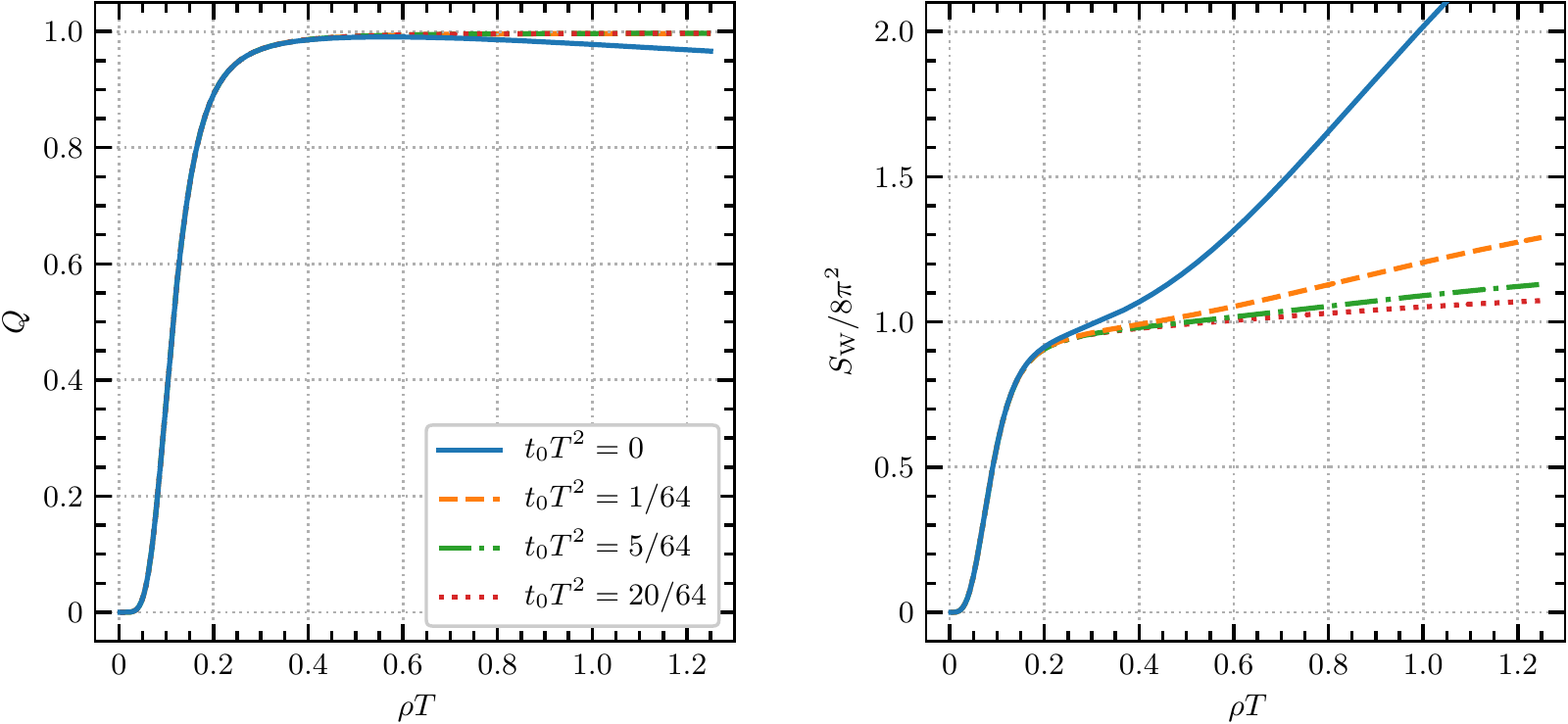}
    \caption{\label{fig:boundaryflow}Caloron topological charge (left) and Wilson action
      (right) for different values of $t_0T^2$ as defined
      in \eqref{eq:boundaryflowtime} on an $8 \times 48^3$ lattice.}
\end{figure*}

In writing down caloron field configurations, we
cannot avoid the discontinuity at the spatial boundaries of our
box. We will try to minimize
the damage by smearing out the boundary
discontinuities with gradient flow. Consider the caloron
placed at the center of the lattice and denote the lattice spatial
extent as $L$. We then flow the links using a flow time that depends
on the relative distance
\begin{equation}
    d = \sqrt{\left(x-z\right)^2}
\end{equation}
of the base point of the link $U_\mu(x)$ and the center of the instanton. The flow time depth gets modified as
\begin{equation}
    t(d) = \begin{cases} 0, \hfill d < \frac{L}{4}
    \\
    \frac{t_0}{2} \left( 1 + \sin \left[ \frac{4\pi}{L} \left( d - \frac{3}{8} L \right) \right] \right),
    \hfill \frac{L}{4}\leq d\leq\frac{L}{2}
    \\
    t_0, \hfill d > \frac{L}{2} \end{cases}
\label{eq:boundaryflowtime}
\end{equation}
where $t_0$ is the ``normal'' flow time. This is nothing but a smooth
interpolation between zero flow (close to the center of the caloron)
and full flow (close to the boundary). With this procedure we reduce
boundary effects while the core of the caloron remains unaffected. In
Fig.~\ref{fig:boundaryflow} we show both the topological charge and
the Wilson action of the caloron for different values of $t_0$. We
observe that the Wilson action suffers significantly more from
boundary effects than the topological charge. Applying this modified
version of Wilson flow indeed reduces boundary effects. We find that a
flow time of $t_0 T^2=5/N^2_\tau$ is sufficient to satisfactorily reduce most
of the boundary effects.

\bibliographystyle{JHEP}
\bibliography{refs}

\end{document}